\documentclass[fleqn,10pt]{wlscirep}
\usepackage[utf8]{inputenc}
\usepackage[T1]{fontenc}
\usepackage{fancyhdr}
\usepackage[onehalfspacing]{setspace}
\title{Real or Virtual? Using Brain Activity Patterns to differentiate Attended Targets during Augmented Reality Scenarios}

\author[1,*]{Lisa-Marie Vortmann}
\author[1]{Leonid Schwenke}
\author[1]{Felix Putze}
\affil[1]{Cognitive Systems Lab, Department of Mathematics and Computer Science, University of Bremen, Bremen, Germany}

\affil[*]{vortmann@uni-bremen.de}

\begin{abstract}

  Augmented Reality is the fusion of virtual components and our real surroundings. The simultaneous visibility of generated and natural objects often requires users to direct their selective attention to a specific target that is either real or virtual. In this study, we investigated whether this target is real or virtual by using machine learning techniques to classify electroencephalographic (EEG) data collected in Augmented Reality scenarios. A shallow convolutional neural net classified 3 second data windows from 20 participants in a person-dependent manner with an average accuracy above 70\% if the testing data and training data came from different trials. Person-independent classification was possible above chance level for 6 out of 20 participants. Thus, the reliability of such a Brain-Computer Interface is high enough for it to be treated as a useful input mechanism for Augmented Reality applications.\\
  

\end{abstract}
\begin{document}

\flushbottom
\maketitle
%
%
\thispagestyle{fancy}

\section{Introduction}
One of the many challenges that our brain is faced with daily, is the filtering and processing of vast amounts of information about our surroundings. The input recorded by our auditory, visual, olfactory, gustatory, proprioceptive, and tactile senses is immense at almost any given moment. To survive in a world of sensory overload, we need to give a meaning to this available information and focus on the most important aspects of the input. The cognitive process of directing this focus on a selected sensation is the core of attention mechanisms \citep{Kinsbourne}. Sublte differences can still be found between different definitions of attention because many processes are still under examination and the meaning, assumptions, and implications about the importance of consciousness, concentration, willingness, allocation of resources, memory, and vigilance are yet to be understood. \\
Augmented Reality (AR) is a relatively new type of user interface that combines real and virtual content. At its core, it is the display of generated information and objects into a natural environment. This merges the near-infinite memory capacity and processing power of computers with human intelligence, information processing, reasoning, and bodily adaptability. The presentation of the virtual content can happen through hand-held devices or head-mounted displays (HMD) and either be see-through for real surroundings or display a video of the environment with the added virtual content (see Section \ref{sec:ar}). This melting of real and virtual information adds to the sensory input and requires sophisticated capabilities to retain attention in order to avoid distraction. \\
Visual attention refers to the conscious and unconscious filtering and selection of visual input \citep{Rensink}. In many cases, this process is linked to gaze behavior, assuming that we direct our eyes at the attended targets (overt attention). Intuitively, the analysis of eye tracking data is often chosen as a means for attention detection. However, the exact gaze point detection requires constant recalibrations to correct for slight movements and sometimes, eye tracking data cannot be recorded with sufficient accuracy if a participant is wearing glasses or has another eye condition. Also, attention is more complex than just target-orientation. Different aspects and forms of attention (for example internal and external visual attention, split attention, or overt and covert visual attention) cause different neural activity patterns in the brain \citep{Kinsbourne}. Therefore, brain imaging techniques can be used to study and detect complex attention mechanisms. \\
Electroencephalography (EEG) is one technique that is used to measure such brain wave patterns. The electrical activity is measured with electrodes that are placed on the scalp and recorded by a computer. With its high temporal resolution and the possibility to use a mobile setup, EEG is a good solution for real-time brain-computer interfaces (BCI) and cognitive state classification in general. \\
Especially if the attentional targets have certain, distinct properties, brain pattern analysis is reliable to classify the current targets of attention, independent of gaze. For example, when the luminance of an attended target flickers in a steady frequency, the same frequency can be observed as a neural response in the brain. This phenomenon is called steady-state visually evoked potential (SSVEP) and it is a robust detection mechanism for visual attention independent of eye movement to the target and even possible for peripheral vision \citep{SSVEP}.\\ 
Augmented Reality introduces two types of distinguishable attention targets in one scene: real ones and virtual ones. This distinction did not have to be made before, because it is unique to AR but offers interesting information about how users interact with AR and how they process available information. Some AR applications might profit from the information whether the user's attention is on real or virtual information and adapt their user interface to the prediction. For example, an application that guides a worker while building a complex machine should not show big virtual objects and explanation if the worker is currently focusing on the real object. The future goal is to work towards such a real-time classifier of attention on real and virtual objects that could support pure eye tracking results for attention direction and supply the application with information about the current attention of the user. Depending on the context, as seen in the example, an interface adaptation might be appropriate to reduce unwanted distractions and improve the usability of AR.\\

In this work, we performed a study to test how well we can classify attention on real and virtual objects in a controlled Augmented Reality setting based on EEG data. We implemented a pairs game that has a virtual and a real set of cards and recorded 20 participants during phases of remembering. The collected EEG data was classified using a shallow convolutional neural net that was built analogously to a filter bank common spatial patterns feature extraction approach. We compare the results to a simple eye tracking classification approach and test generalizability by analyzing the spectral density of the EEG data for each participant and test a person-independent classifier. \\

\subsection{Augmented Reality Technology}\label{sec:ar}
\begin{figure}
    \centering
    \includegraphics[width=0.5\textwidth]{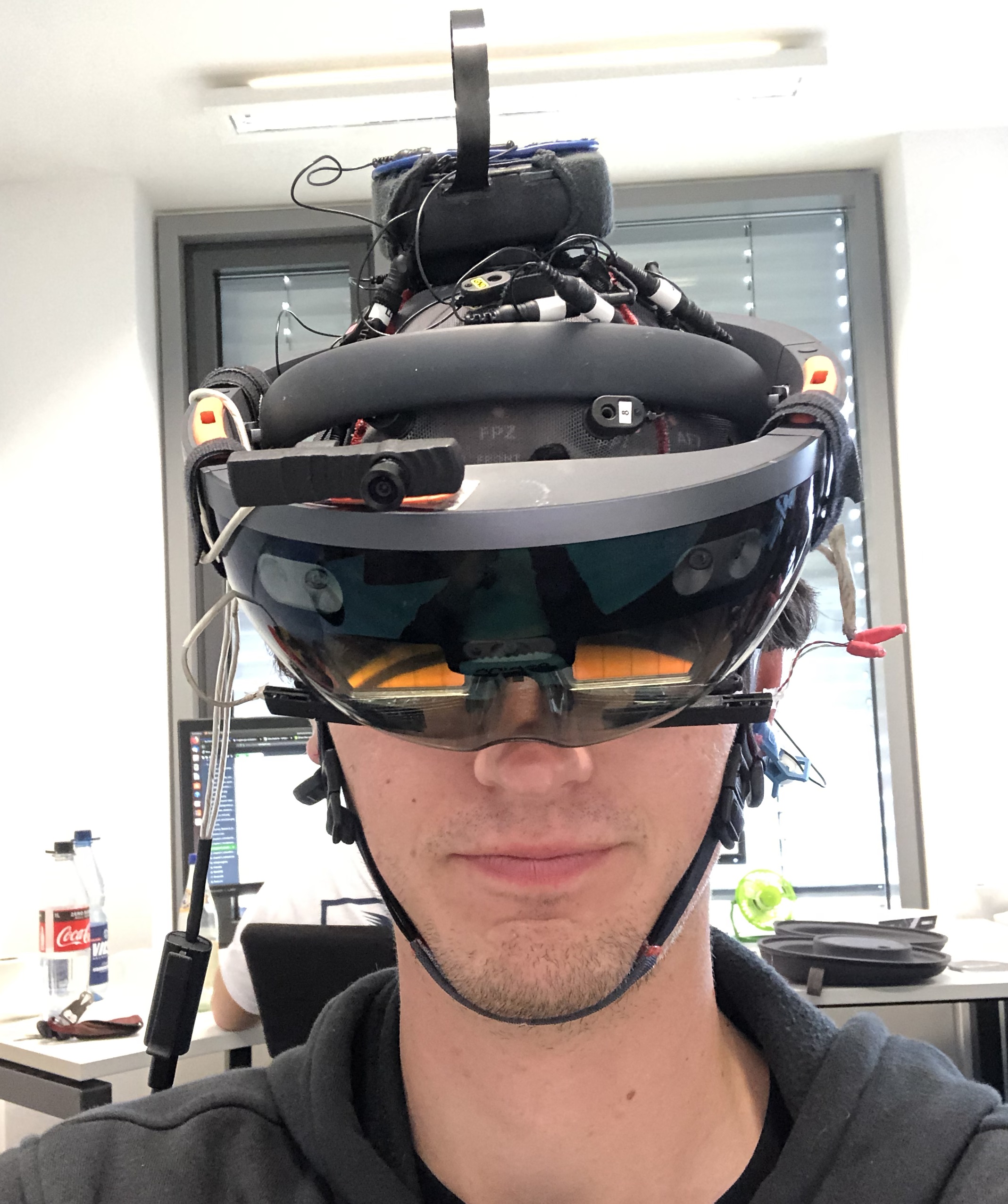}
    \caption{Setup of the EEG and the HoloLens during the experimental session}
    \label{fig:setup}
\end{figure}

According to \cite{Azuma}, three main characteristics define an AR system: (1) It combines real and virtual content, (2) the interaction with the system happens in real-time, and (3) a reaction and three-dimensional integration of virtual content in the real surroundings takes place. As mentioned before, AR interfaces can be realized through different devices which \cite{simohammed2017} categorize into three types: \textit{Video See-Through Augmented Reality devices} record their surroundings and display them through a video with the added virtual content. This version is often used for AR applications on mobile phones and tablets. \textit{Optical See-Through Augmented Reality devices} instead, use a transparent screen that only displays the virtual content while allowing the user to still see the real surroundings. In \textit{Projective AR}, the virtual content is projected onto real objects in the environment. \\
In this work, we use an Optical See-Through AR device with an HMD that was developed by Microsoft (HoloLens Gen 1, see Figure \ref{fig:setup}) because, at the time of the study, it is one of the most advanced devices. All the virtual information is directly projected into the participant's field of view, several cameras scan the surroundings for correct object placement, and interaction with the system is possible through voice control, gestures, and external clicking devices. The participant movement is tracked in relation to the real world, in order to stick virtual objects to real places, even if they are outside of the field of view.\\
Currently, AR technology still has its limitations (frame rate, resolution, projection area) and users are usually able to distinguish between real and virtual objects. First of all, virtual objects are usually brighter because they originate from a light source within the device. Additionally, augmented content can withstand physical laws. While the virtual content can be influenced by the real content, the real content can never truly be influenced or changed by virtual content and this flaw can become obvious in different scenarios (i.e. object movements). Another reported projection flaw is that objects can appear to be floating in times when they should not. Further limitations are caused by the 2.4 megapixel display of the device. The 16:9 ratio of the screen offers an HD resolution of 1280 × 720 pixels per eye but restricts the natural field of view. Thus, a virtual object may disappear from the screen, while the natural surrounding where it was placed is still visible for the user.\footnote{Source: https://docs.microsoft.com/en-us/hololens/hololens1-hardware} The display is updated with a refresh rate of 60 Hz.

\subsection{Related Work}\label{sec:related}
To the best of our knowledge, there are no scientific publications that dealt with brain activity pattern differences for virtual and real objects during focused attention in AR. However, the general combination of Augmented Reality and Brain-Computer Interfaces has been of high interest lately. \cite{simohammed2017} published a state-of-the-art summary for AR and BCI and their scope of application. In their work, recent improvements in the combination of both technologies become apparent, however, they also critically reflect on problems and shortcomings, such as setup and movement artifacts. The problematic combination of two head-mounted systems, like EEG and Augmented Reality headsets, has also been mentioned by  \cite{putze2019}. Many AR-BCIs or Brain-controlled AR systems make use of specific neural responses such as SSVEP or the P300. These will not be discussed further for this work because we will not work with event-related or steady-state evoked potentials because they are only applicable in specific AR scenarios and applications. \\
The general use of EEG recordings to assess visual attention has been studied and was proven to be successful in many studies \citep{Liang, krigolson, busch, Sauseng}.\\
There are several studies that have focused on the modeling of cognitive states for EEG-based BCI systems. For example, these BCIs can differentiate between types of attention, mental workload, tiredness, or emotions \citep{myrden}. \cite{zhang} used EEG data to estimate two states of mental fatigue on a single-trial basis with an accuracy of 91\%. To locate user attention and reduce mental workload, \cite{putze2013} used eye tracking and EEG data for spatio-temporal event selection. They achieved 91\% temporal and 86\% spatial accuracy for a static paradigm. A mobile setup was tested in \cite{liu2013}, where students attentiveness was measured in a classroom and correctly classified with an accuracy of 76.82\%.\\
Such passive BCIs \citep{zander} have also been combined with Augmented Reality in several scenarios.
 \cite{barresi2015} claimed that BCI-checked surgical training for users is better than normal training. They combined a BCI with AR and estimated the level of attention. In \cite{vortmann2019}, the authors showed that it was possible to classify internal and external user attention in an Augmented Reality paradigm and in \cite{vortmannchi}, a first real-time attention-aware smart-home system in AR was implemented and tested. It was shown that the usability was improved and the distraction decreased by including system behavior restrictions based on the detected internal or external attention of the user. As input, EEG and eye tracking data was used.\\

Many BCI systems use gaze as an active input mechanism for controlling an application \citep{silbert2000}. However, there is more information in the viewing behavior than only the current gaze point. It has been shown that it is possible to classify emotions \citep{florea2013}, mental states \citep{day1964}, cognitive deficits \citep{hutton2006}, and internal thoughts \citep{Huang} from eye gaze. 

\subsection{Hypotheses}
Based on the related work, the current quality of augmented content and knowledge about neural processing of visual information, we hypothesize that activity patterns in the human brain are different for visual attention of real-world objects and virtual objects in Augmented Reality. This hypothesis is based on the fact that the virtual content is still recognizable as such by the user. Thus, it should evoke a noticeably different neural response. Building up on the assumption that there is a detectable difference in the neural response, we hypothesize that state-of-the-art machine learning algorithms should be able to learn this difference and build models for both cases of attentional focus (real and virtual object). Precisely, our main hypothesis is stated as follows:
\begin{itemize}
    \item[H1:] In a controlled Augmented Reality environment, a person-dependent EEG data classification of real or virtual visual attention is possible with an accuracy significantly higher than chance level. 
\end{itemize}
One major discussion point that is often critical of EEG-based attention classification is the fact that eye tracking data is easier to collect and leads to even better results. Newest Augmented Reality headsets are even supplied with a built-in eye tracker (i.e. Microsoft's HoloLens 2). This method reaches its limitations for small or overlapping content and is highly dependent on an accurate eye tracker calibration if the gaze point is used to define the current attentional target. Slight movements of the eye tracking device in relation to the eyes would influence the result and constant recalibration would be necessary. Instead, the gaze patterns can be analyzed for differences, as it is often done in eye tracking studies on other cognitive phenomena (see Section \ref{sec:related}). In our scenario, the viewing behavior for virtual and real objects would have to be noticeably different. We assume, that there are only marginal differences and thus, we hypothesize additionally: 
\begin{itemize}
    \item[H2:] In a controlled Augmented Reality environment, basic person-dependent eye tracking data classification of real or virtual visual attention is not significantly more reliable than the person-dependent classification of simultaneously recorded EEG data. 
\end{itemize}
Assuming that the distinguishable neural activity patterns are evoked by the nature of the virtual representations, they should be similar across participants. Cross-participant EEG pattern recognition for person-independent Brain-Computer Interfaces has been a challenging, but desired topic in the field. Individual differences among the participants and users lead to lower classification accuracies compared to models that were trained on person-dependent data. However, we want to analyze whether the models still generalize over participants. 
We formulate our third hypothesis as follows: 
\begin{itemize}
    \item[H3:] In a controlled Augmented Reality environment, a classifier trained on person-independent EEG data can predict real or virtual visual attention of a new participant with an accuracy significantly higher than chance level. 
\end{itemize}

The focus of this study lies on H1, with H2 and H3 being supporting hypotheses. H2 is taken as a motivation to study this topic and H3 is a preliminary analysis to inspire further thoughts in the direction of training-free real-time BCIs.
\begin{figure}

\includegraphics[width= \textwidth]{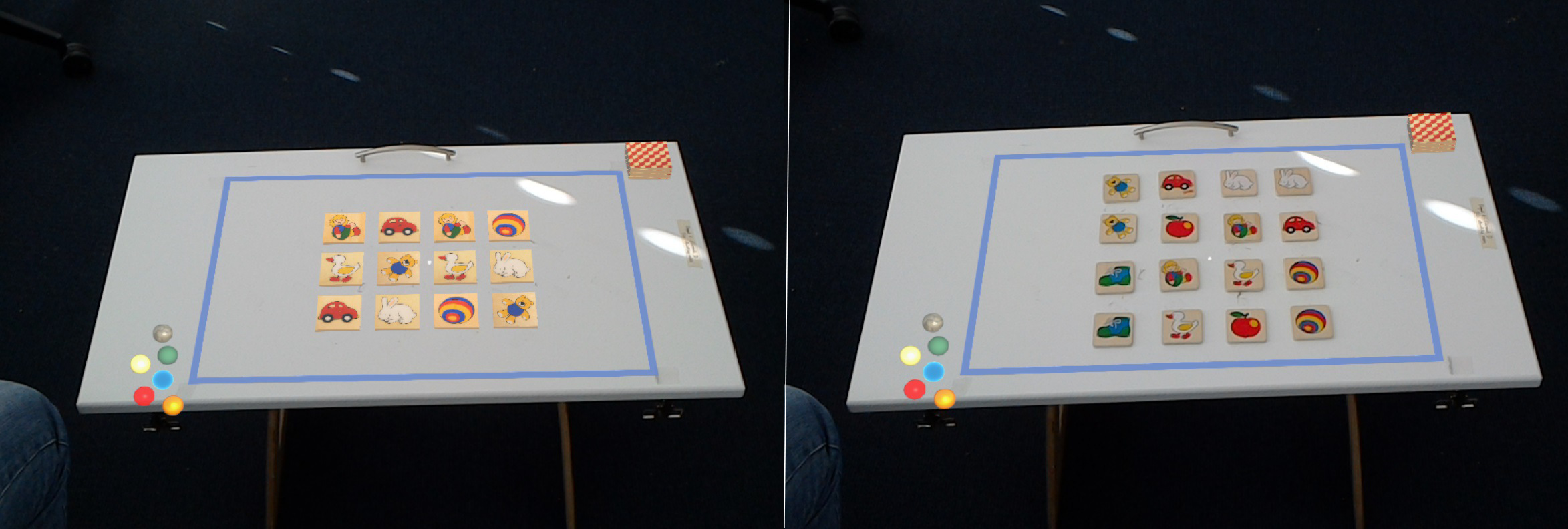}
\caption{Participant's view of the two task conditions. Left: "Real" condition with 4x4 field size; Right: "Virtual" Condition with 4x3 field size; Both:  Virtual field size border, marbles and deck of cards  }
\label{fig:Screen}
\end{figure}
\section{Experimental Design}\label{sec:design}
The major requirements for the experimental task were (1) to ensure retained attention on a real or virtual object over a controlled period of time, (2) a high similarity between the real and virtual trials and their objects beyond the mode of presenting the scene elements, and (3) to avoid strong artifacts in the data. To avoid these artifacts (i.e. caused by movements), we decided to use a controlled and static setting. The retained attention on specific parts of the field of view was achieved by directing attention through gamification of the task. We implemented a card game that follows the idea of the popular game PAIRS. The goal was to remember the positions of matching pairs of cards and recall them. The implemented task consisted of two conditions: "Real" and "Virtual". In general, some virtual elements were present in both conditions to optimally simulate a typical Augmented Reality scenario (the border of the field, marbles by the side, a deck of cards). The crucial difference between the two conditions was that in the "real" condition, the pairs cards that contained the objects were real physical cards, whereas, in the "virtual" condition, the cards were virtual objects that were displayed by the Augmented Reality device. The virtual and real cards were identical in size and displayed the same pairs of matching figures. The playground was a light 90cm x 40cm  wooden board placed in the same height as the person was sitting. This placement was chosen to give the participant a good overview of the full set of cards. Field sizes were chosen randomly for both conditions to vary the difficulty. The possible fields were 5x5, 4x3, 7x2, and 4x4 (see Figure \ref{fig:Screen}).\\

Because the interaction between a participant and the real or the virtual cards would not be identical due to limited input mechanisms for AR, we decided to play the traditional pairs game with adjusted rules in a 1-person format, without an opponent. In the beginning, the cards are presented with their symbols facing upwards. The participant can look at the cards in their positions for 20 seconds and is asked to remember the positions of as many pairs as possible. This phase is called Memory-Phase. Afterward, in the Recall-Phase, the cards are turned around and the participant has 20 seconds to find as many correct pairs as possible by selecting them (in the "Virtual" condition) or picking them up (in the "Real" condition) one by one to turn them around, as typical in the traditional pairs game. If the participant makes a mistake, the wrong cards are turned back around but the Recall-Phase continues. In the end, feedback is given about the correct number of pairs that were detected. All participants were asked not to guess randomly but focus on the cards on the Memory-Phase.\\
The only data that was used in the analysis, was the data that was recorded during the Memory-Phase. In this phase, the participants retained their attention either on only real or only virtual content in an Augmented Reality scenario, while no other actions or cognitive tasks could make a difference between real and virtual trials. This ensures that any difference which is found by the classifier can be tied to the difference in attention on virtual vs. real objects. The data from the Recall-Phase was not used, because during the recall the participant had to perform the action of turning the cards around. This was done manually in the "Real" condition. In the "Virtual" condition, the participant used the visual "gaze point" that is visible through the Augmented Reality device and a Bluetooth clicker for selection. Since we did not want these differences to be learned by the model, we did not use the data. However, we played the full game to encourage the motivation of the participants and to record the number of correct pairs as a measure of how well the Memory-Phase was performed by the participants. \\

\begin{figure}
    \centering
    \includegraphics[width=0.8\textwidth]{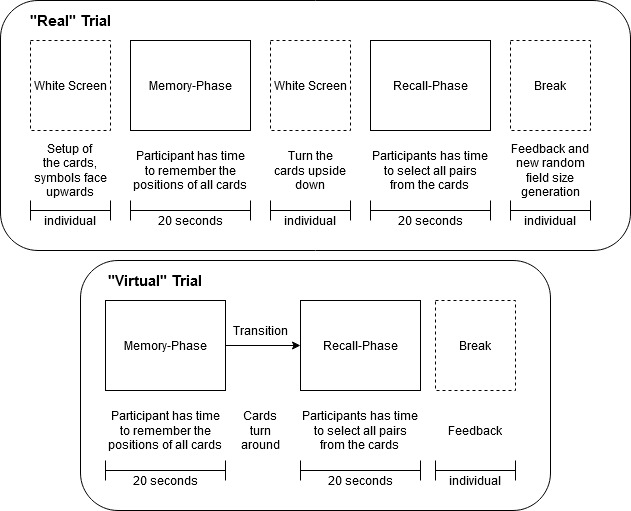}
    \caption{Step-by-step procedure of the two task conditions with times and performed actions}
    \label{fig:ablaub}
\end{figure}

One advantage for the player in the "Virtual" condition was that all cards were simultaneously presented to start the Memory-Phase and simultaneously turned over after the Memory-Phase. In the "Real" condition, this was not possible because the cards had to be laid out and turned around by hand by the experimenter. The sharp beginning and cutoff of the "Virtual" condition were simulated by a white screen covering the whole visual field of the Augmented Reality device. This time period was not used during the classification process. Because of the lighting conditions, the white screen covered all the cards so that they would not be visible, while the experimenter prepared the field. The exact procedure of each condition of the task can be seen in Figure \ref{fig:ablaub}.\\

The game was implemented in Unity (version 2018.4.2f1) using the HoloToolKit (version 2017.4.3.0) for compatibility with the Microsoft HoloLens Generation 1.

\section{Methods}
For this study, we recorded data from 20 participants with normal or corrected to normal vision (age 26.1 $\pm$ 7.2; 7 female). 9 of the participants were university students at the time of the recordings (7 in the field of computer science). In a preliminary questionnaire, 11 participants reported that they had experiences with Virtual Reality and 7 participants reported that they had used an Augmented Reality device before. After an introduction to the experiment, all participants gave written informed consent to the recording and storing of their data in a completely anonymized fashion. The study was approved by the local ethics committee.

\subsection{Experiment Session}
All experiment sessions took place in the same room of an office building. The room was quiet but not shielded and both sunlight and artificial light were present. A less controlled experiment environment was chosen for a justified comparison to possible real-life applications. The same experimenter attended and instructed all sessions to avoid differences between performance and explanations during the trials. All participants were instructed to come without makeup for better eye tracking results and to not wash their hair on the day of the experiment, for a higher EEG calibration accuracy. The chair and playing field were set up before the session. \\
In the beginning, the participants filled out a demographic questionnaire and received a written explanation of the task to ensure that no information was left out. As the first step, the EEG cap was set up and the HoloLens with the eye tracker was adjusted on top of the cap (see Figure \ref{fig:setup}). For details on the EEG and eye tracking, see Section \ref{sec:datarec}. After the setup was completed and the participant was seated in the experiment chair, the eye tracker was calibrated using the manual 3D-marker calibration of Pupil-Labs software Pupil Capture\footnote{https://docs.pupil-labs.com/core/software/pupil-capture/\#calibration-methods}. Unfortunately, eye tracking recordings are only available for 13 participants because of technical problems. All comparisons between eye tracking and EEG data in this study will be based on the 13 participants only.\\
The experiment was controlled via an experiment terminal by the experimenter. Before the main trials started, each participant executed tutorial trials to get accustomed to the operation of the game. The number of tutorial trials was chosen individually, depending on the feedback of the participant. In total, each participant completed 20 trials of the "real" condition and 20 trials of the "virtual" condition, resulting in a total of 40 trials. These trials were performed in blocks of four trials of the same condition. After each block, the condition of the next block was generated randomly with the constraint, that a maximum number of 5 blocks per condition was possible. The block design was chosen to facilitate and shorten the experimental setup during trials, especially for the "real" trials, where the experimenter was required to set the cards up manually. The field size was also generated randomly for each trial. During the "virtual" trials, errors and correctly recalled pairs were recorded by the application, while for "real" trials, the experimenter noted performance results by hand.\\
On average, the participants spent less than 60 minutes with the performance of the experiment. Afterward, all participants answered a questionnaire regarding their perception of the task. The questions compared the perceived difficulty, interaction, and usability of both conditions. Additional free feedback was collected.\\
Trials, during which technical or environmental problems occurred, were excluded for the analysis. This led to a reduced number of trials for participants 5,6,7,8, and 10 and an overall average of 0.5 deleted trails. The reduced trial number is corrected for in all statistical analyses. 

\subsection{Data Recording}\label{sec:datarec}
During the main trials, we recorded EEG, eye tracking, and task data with matched timestamps using the Lab Streaming Layer system (LSL\footnote{https://github.com/sccn/labstreaminglayer}). The task data included the beginning and end of trials, as well as their condition, and the results of the recall phase. This data was used for windowing and performance analysis. For the communication between the HoloLens and LSL, we used LSLHoloBridge.\footnote{https://gitlab.csl.uni-bremen.de/fkroll/LSLHoloBridge} For details on the architecture, see  \cite{vortmann2019}.\\

The EEG data was recorded using a wireless g.Nautilus EEG-headset from g.tec\footnote{https://www.gtec.at/product/gnautilus-research/} and 16 gel-based electrodes. The positions of the electrodes are based on the 10-20-system, covering the whole scalp, but were adjusted to have minimal interference with the placement of the HoloLens. This resulted in the following placement: Cz, Fp2, F3, Fz, F4, FT7, C3, Fp1, C4, FT8, P3, Pz, P4, PO7, PO8, and Oz. We used a 500 Hz sampling rate during the recordings and impedances were kept below 30 k$\Omega$. The data was recorded with an ear-reference and AFz as the ground electrode. Electrodes with insufficient data quality after manual inspection during the calibration were excluded during the analysis (on average 1.85 $\pm$ 1.66 electrodes).\\

Since there is no integrated eye tracker in the HoloLens Generation 1, the PupilLabs binocular eye tracker was mounted to the device. The two cameras that record the eyes are placed under the screen of the head-mounted display and the world-camera is fixated above the screen. The cameras record the eyes with a sampling rate of 120 Hz. Pupil Capture was used to record the gaze position and pupil diameter of the participant. We decided to use basic 2D gaze point coordinates for the analysis in this study, to reflect the eye tracking capabilities of the built-in eye tracker of the HoloLens Generation 2 (as of June 2020). \\ The pupil diameter reacts to the brightness of the surroundings. Thus, we decided to disregard it for the analysis, because for a generalized claim about Augmented Reality scenarios, the changing brightness will depend highly on the environment and task and will not be as stable as in our controlled setting. The lighting conditions between the recordings of participants may vary. However, since the eye tracking data is used for person-dependent analysis, this will not be discussed. As reported before, technical difficulties resulted in a reduced set of eye tracking data from only 13 of 20 participants. The average eye tracking accuracy after the calibration was 2.49 $\pm$ 0.51 degrees.

\subsection{Analysis}\label{sec:analysis}
The data analysis, including the preprocessing and the classification, was performed with Python 3.6. All preprocessing steps were kept to a minimum, aiming at a feasible pipeline for a real-time BCI.\\
Error statistics of the participants were collected and used to test for the desired focused attention during the task. A very low performance would would have resulted in the exclusion of the participant. However, that did not happen, and single trials with bad performance results were not excluded.\\

The EEG data was preprocessed using the MNE toolbox.\footnote{https://mne.tools/stable/index.html} After the channels with insufficient were excluded, the data was band-pass filtered between 3 and 45 Hz using a windowed FIR-filter, and an additional notch-filter at 50 Hz (power source frequency) was applied. All excluded channels were interpolated and the data was re-referenced to average reference. We did not visually inspect the EEG recordings for artifacts, neither were any artifacts cleaned automatically. Again, this approach was chosen with a real-time BCI in mind.\\
Based on the experimental markers, 3 second EEG windows were extracted. This window length was chosen because it allows claims about the feasibility of a real-time BCI in this context. As mentioned in Section \ref{sec:design}, only the data from the Memory-Phase will be used for classifier training and testing. The 20-second Memory-Phase was cut into 5 non-overlapping windows (3-6,6-9,9-12,12-15, and 15-18 seconds after Memory-Phase onset). The first and last seconds of each Phase were left out because the probability for artifacts and missing attentional focus is higher.\\
The cleaned, windowed raw EEG data was used as the input for a shallow convolutional neural network. The network was tested and implemented by \cite{Schirrmeister} and is built following a Filter Bank Common Spatial Pattern (FBCSP) feature extraction pipeline. Based on previous experiments, the learning rate of the model was adjusted to 0.0015 with a weight decay of 0. Their suggested cropped training approach was applied with automatic settings. In all analyses of this study, the neural net was trained for 150 epochs.\\
\subsubsection{Trial-Oblivious Approach}
For the person-dependent analysis, we first tried a randomized, stratified training-testing split with 30\% testing data and repeated the training and testing with 10 random splits for each participant for a better accuracy estimation. The splits of the epochs were independent of the trials they belonged to, thus \textbf{trial-oblivious}.\\
\subsubsection{Trial-Sensitive Approach}
Since we always extract 5 data windows from each trial, the effect of belonging to the same trial within the recording might play an important role for the model. To test and correct for this effect, we additionally used a training-testing split that was trial sensitive. For this approach, all windows from the same trial were either in the training or in the testing data. This \textbf{trial-sensitive} split was also performed ten times, randomly but stratified, with 30\% testing data. \\
In the next step, we analyzed the accuracy that was achieved during the trial-sensitive randomized approach, based on the position of the extracted window within the trial. The question to be answered was, whether any time-frame of the 20-second trial achieved significantly lower or higher classification accuracies than the other extracted windows.\\
\subsubsection{BCI-Approach}
As the last method for splitting into training and testing data, we chose an approach that represents the real application of a BCI. In the \textbf{BCI-approach}, we maintained the chronological order of trials, i.e. trained the model on the windows of the first trials for each condition and tested the model performance on the windows of the last trials of each condition (70:30 split). If this classifier were used in a real-time BCI, the classifier would be trained on training data that is collected and labeled in a controlled setting before the classifier is used for testing trials in the application.\\

\subsubsection{Eye Tracking}
For the \textbf{eye tracking classification}, the same windowing was chosen as described for the EEG data. We followed the feature extraction and classification procedure as described in \cite{vortmannicmi}, using the scikit-learn toolbox \citep{scikit}. The calculated features were based on the recorded x and y gaze point coordinates and included information about the outlier quote, fixations, saccades and gaze velocity and distance. Again, the training and testing split was performed trial-sensitive but randomized 10 times in a stratified manner. The reported accuracies are the average over all ten fitting and testing runs of the LDA. The NN could not be used for the eye tracking data because it was specifically designed for eye tracking data. The design and evaluation of suitable Neural Nets for the eye tracking classification in the presented task are not within the scope of this work. \\
We were also interested in the combination of both modalities. In a late fusion approach, we calculated the average accuracy over ten runs for each participant when the EEG prediction was only used if the confidence that was estimated by the NN surpassed a fixed threshold. In all other cases, the eye tracking prediction was taken. 
\begin{table}[]
\begin{tabular}{l||l|l|l|l}
Approach           & Data & Train:Test  & Split-Restriction                                                                                                                          & Classifier \\ \hline \hline
Trial-Oblivious    & EEG  & 70:30       & Stratified  & NN \\ \hline
Trial-Sensitive    & EEG  & 70:30       & \begin{tabular}[c]{@{}l@{}}Windows from the same trial are either\\ all in the training set or all in the test set,\\ Stratified\end{tabular} & NN \\ \hline
BCI-Approach       & EEG  & 70:30       & \begin{tabular}[c]{@{}l@{}} First 70\% of the trials of each label are \\ in the training set, last 30\% of the trials\\ of each label are in the test set          \end{tabular}        & NN \\\hline
Eye Tracking       & ET   & 70:30       & \begin{tabular}[c]{@{}l@{}}Windows from the same trial are either\\ all in the training set or all in the test set,\\ Stratified\end{tabular} & LDA        \\ \hline
Late Fusion     & \begin{tabular}[c]{@{}l@{}}ET \\ EEG\end{tabular}   & 70:30       & \begin{tabular}[c]{@{}l@{}}Windows from the same trial are either\\ all in the training set or all in the test set,\\ Stratified\end{tabular} & \begin{tabular}[c]{@{}l@{}}NN\\ Threshold\\LDA\end{tabular}       \\ \hline
\begin{tabular}[c]{@{}l@{}}Person-\\Independent\end{tabular} & EEG  & Leave-1-out & \begin{tabular}[c]{@{}l@{}} No data of the test subject is in the \\ training set       \end{tabular}   & NN
\end{tabular}
\caption{Description of all performed classification accuracy analyses}
\end{table}
\subsubsection{Person-Independent Approach}
For the third hypothesis, we tested whether the classification of EEG data for this task is possible above chance level for a \textbf{person-independent classifier}. This means that the classifier is never trained on data from the participant who's data is to be classified. The same preprocessed EEG data was taken from all participants and the same neural net as described before was trained in a leave-1-participant-out fashion for all participants. Thus, the model was trained on the data of 19 participants and tested on the remaining full data set of one participant. To add to the understanding of generalizability of the differences in the EEG data, we computed the mean Power Spectral Densities (PSD) for the Alpha (8-14 Hz), Beta (14-30 Hz), Gamma (30-45 Hz) and Theta-band (4-8 Hz) for each electrode of each participant. This results in \textit{16 electrodes x 4 frequency bands = 64 features} that were compared for significant differences. We used MNEs Welch-method to calculate the PSDs. The results for each window were scaled between 0 and 1 based on the minimum and maximum for each feature of each participant individually before computing whether there are significant differences between the "Real" and "Virtual" condition if pooled over all subjects. For this analysis, a significance level of $\alpha= 0.001$ was chosen, for a meaningful statement despite the high number of available data windows (n = 3940, approximately 200 windows per subject).
\subsubsection{Evaluation}
We evaluated the accuracy, precision, recall, and F1-score for all approaches but we will base our discussion of the results on the accuracy of the classifier. Due to the balanced class distribution, the chancel level for correct window classification is 0.5. Thus, accuracy should represent the quality of the classification well.\\

In order to rate whether the classification accuracy is significantly higher than chance level, an approach suggested by \cite{mullerputz} was followed. Based on the assumption that this two-class paradigm follows a binomial distribution with \textit{n = number of test trials} and \textit{p = 0.5}, we can assume that the confidence intervals are given by
\[p \pm \sqrt{\frac{p(1-p)}{n+4}} z_{1-\frac{\alpha}{2}} \]
with $z_{1-\frac{\alpha}{2}}$ being the $1-\frac{\alpha}{2}$ quantile of the Normal Distribution with a mean of 0 and a standard deviation of 1. The upper border of the interval is calculated for the claim "better than random". If not stated otherwise, we chose the significance level of $\alpha = 0.05$ for all statistical test in this study.\\

\section{Results}
After the evaluation of the performance results, no participant had to be excluded from the analysis. All participants detected more correct pairs than incorrect pairs on average. Summarized over all performed trials by all participants, 5.22 pairs were detected correctly and only 1.5 errors were made. A difference between the conditions can be observed: 77.34\% of the trials in the "virtual" condition were complete within the time limit with all pairs, whereas this was the case in 81.35\% of the "Real" trials. However, due to the different procedures during both conditions, with longer pauses in the "Real" condition and a higher technical difficulty in the "Virtual" condition, no conclusions should be drawn from these results. Importantly, the performance was high enough to assume focused attention during the Memory-Phases.\\
The results of the questionnaires and individual comments also did not lead to any reasonable exclusion of trials or participants. Overall, the interaction with both conditions was comfortable. It was reported, that the Recall-Phase was harder to perform in the "virtual" condition, but this has no effect on the data of the Memory-Phase. Additionally, the questionnaire results show, that the memorization of the cards was neither too hard, nor too easy with an average score of 2.2 $\pm$ 0.95 for the "Virtual" condition and 2.55 $\pm$ 1.1 for the "Real" condition on a scale from 1 = easy to 5 = hard.
12 participants rated the "Virtual" condition as preferable and more fun. There was no correlation between classification results and experience with AR.
\begin{table}[h]
\begin{tabular}{l|cccc}
Participant & Trial-Oblivious & Trial-Sensitive & BCI-Approach & Person-Independent \\\hline
1           & 0.96*   & 0.69*          & 0.55         & 0.48               \\
2           & 0.91*    &0.70*         & 0.63*        & 0.53               \\
3           & 0.97*     & 0.86*      & 0.83*        & 0.50               \\
4           & 0.94*     & 0.79*        & 0.55         & 0.63*              \\
5           & 0.92*     & 0.71*        & 0.65*        & 0.54               \\
6           & 1.00*      & 0.60       & 0.59         & 0.54               \\
7           & 1.00*      & 0.71*       & 0.58         & 0.56               \\
8           & 1.00*     & 0.90*        & 0.89*        & 0.70*              \\
9           & 0.87*     & 0.65*        & 0.49         & 0.49               \\
10          & 0.64*     & 0.76*        & 0.89*        & 0.47               \\
11          & 0.99*      & 0.57       & 0.40         & 0.64*              \\
12          & 0.97*      & 0.65*       & 0.54         & 0.58*              \\
13          & 0.96*      & 0.73*       & 0.80*        & 0.62*              \\
14          & 0.98*     & 0.64*        & 0.54         & 0.48               \\
15          & 0.86*      & 0.71*       & 0.76*        & 0.51               \\
16          & 0.80*     & 0.77*        & 0.69*        & 0.51               \\
17          & 0.97*      & 0.83*       & 0.83*        & 0.59*              \\
18          & 0.97*      & 0.62*       & 0.58         & 0.53               \\
19          & 0.92*      &0.86*       & 0.90*        & 0.53               \\
20          & 0.99*     &0.71*        & 0.63*        & 0.52               \\\hline \hline
Mean        & 0.93*     &0.72*        & 0.66*        & 0.54               \\
Std        & 0.08       & 0.09       & 0.15         & 0.06               \\
n           & 60       & 60         & 60           & 200               
\end{tabular}

\caption{Overview for trial-oblivious, trial-sensitive, BCI, and person-independent training approaches. Rounded results are shown for all participants individually. Trial-oblivious results are the average accuracy over 10 randomized training-test splits. BCI-approach is the result if the classifier is tested on the last 30 percent of windows. Person-independent training was performed on the data of the other 19 subjects. n = number of test trials. *significantly better than random (based on \citep{mullerputz} with n as indicated, p = 0.5 and $\alpha = 0.05$.)\label{tab:overview}}
\end{table}
\subsection{Person-Dependent Classification}
As a significance measurement, the lower border for "significantly better than random" were calculated as described in Section \ref{sec:analysis}. For all participants with a complete EEG dataset, the classification accuracy had to surpass 62.25\% (n=60). For subject 8, seven trials had to be excluded. The results for this subject were significant if they were better than 64\% (n=45).
The results of the trial-oblivious training sessions for each participant can be seen in the first column of Table \ref{tab:overview}. The average classification result of the 10 runs that were performed per person are all significantly better than chance.  The mean classification accuracy over all participants reached 92.92\% $\pm$ 8.41\%. \\

\begin{figure}[h]
    \centering
    \includegraphics[width=0.8\textwidth]{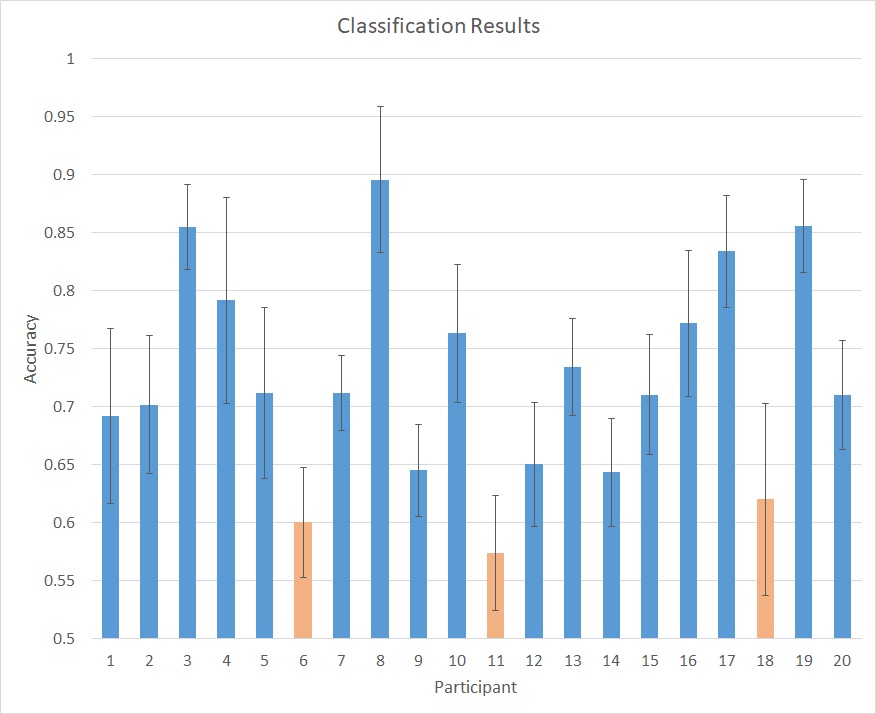}
    \caption{Barplots showing the mean classification accuracy and standard deviation of the trial-sensitive randomized approach performed 10 times for each participant. Orange bars are used for participants, were the average classification accuracy was below the calculated threshold for performance better than random. }
    \label{fig:results}
\end{figure}

In comparison, the trial-sensitive approach reached 72.34\% $\pm$ 8.77\% average accuracy over all participants. We conclude that there is a strong temporal dependency in the data and that the trials-sensitive approach represents the difference between attention on real and virtual objects better. The trial-oblivious approach supposedly overestimates the generalizability of the learned model. We will focus our discussion and further results on the trial-sensitive approach. The classification accuracies for each participant are reported in detail in Figure \ref{fig:results}. From an individual perspective, the classification was better than random for 17 of the 20 participants. We compared the precision, recall and F1 score for both conditions and the results show the same effects as the accuracy measurements, which was expected because of the equal distribution of the two conditions. In further results, the accuracy of the classification will be reported (See Table \ref{tab:results}).  \\

\begin{table}[h]
\begin{tabular}{l|rrr|rrr}
            & \multicolumn{3}{l|}{Real} & \multicolumn{3}{l}{Virtual} \\ \hline
Participant & Precision & Recall & F1   & Precision  & Recall  & F1   \\ \hline
1           & 0.71      & 0.65   & 0.68 & 0.68       & 0.73    & 0.70 \\
2           & 0.67      & 0.81   & 0.73 & 0.76       & 0.60    & 0.67 \\
3           & 0.91      & 0.78   & 0.84 & 0.81       & 0.93    & 0.86 \\
4           & 0.84      & 0.72   & 0.77 & 0.75       & 0.87    & 0.81 \\
5           & 0.72      & 0.69   & 0.71 & 0.70       & 0.73    & 0.72 \\
6\textsuperscript{x} & 0.61      & 0.57   & 0.59 & 0.59       & 0.63    & 0.61 \\
7           & 0.72      & 0.69   & 0.71 & 0.70       & 0.73    & 0.72 \\
8           & 0.85      & 0.94   & 0.89 & 0.94       & 0.86    & 0.90 \\
9           & 0.64      & 0.67   & 0.65 & 0.65       & 0.62    & 0.64 \\
10          & 0.74      & 0.75   & 0.75 & 0.78       & 0.77    & 0.78 \\
11\textsuperscript{x} & 0.57      & 0.61   & 0.59 & 0.58       & 0.53    & 0.56 \\
12          & 0.62      & 0.80   & 0.69 & 0.71       & 0.50    & 0.59 \\
13          & 0.68      & 0.79   & 0.73 & 0.79       & 0.69    & 0.74 \\
14          & 0.63      & 0.68   & 0.65 & 0.65       & 0.61    & 0.63 \\
15          & 0.70      & 0.74   & 0.72 & 0.72       & 0.68    & 0.70 \\
16          & 0.78      & 0.76   & 0.77 & 0.76       & 0.79    & 0.78 \\
17          & 0.79      & 0.87   & 0.83 & 0.88       & 0.81    & 0.84 \\
18\textsuperscript{x} & 0.62      & 0.64   & 0.63 & 0.63       & 0.60    & 0.61 \\
19          & 0.79      & 0.94   & 0.86 & 0.94       & 0.78    & 0.85 \\
20          & 0.72      & 0.69   & 0.70 & 0.70       & 0.73    & 0.72 \\ \hline \hline
Mean        & 0.71      & 0.74   & 0.72 & 0.74       & 0.71    & 0.72
\end{tabular}
\caption{Precision, Recall and F1 score for real and virtual condition of each participant. \textsuperscript{x} = Participants with an average accuracy that was not significant.}
\label{tab:results}
\end{table}
\begin{figure}[h]
    \centering
    \includegraphics[width=0.7\textwidth]{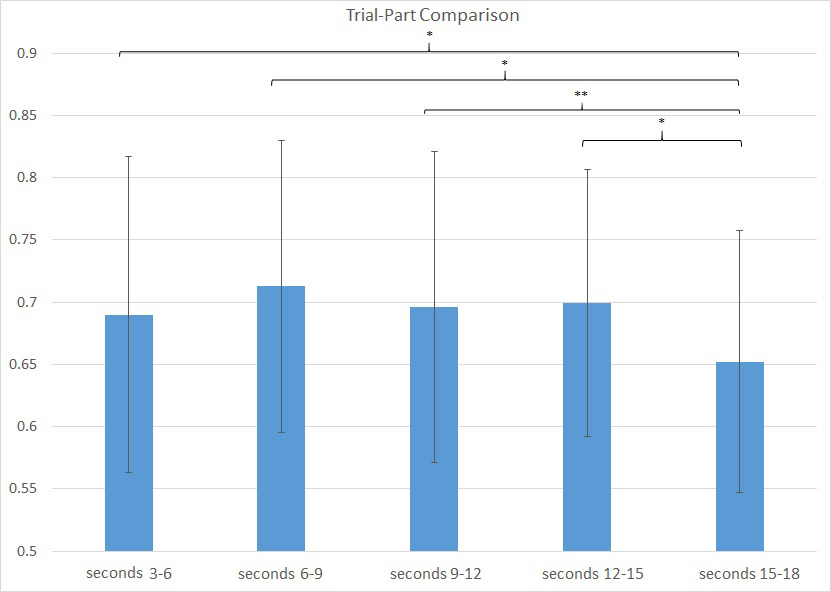}
    \caption{The mean classification accuracy and standard error depending on the timing interval of the extracted window within the trial. The results are calculated on all 10 trial-sensitive randomized runs of all 20 participants. Significant differences between the categories are marked. *Significant ($\alpha < $0.05). **Highly significant ($\alpha < $0.001)}
    \label{fig:part}
\end{figure}

The analysis of the trial-parts showed that the windows extracted from 15-18 seconds after Memory-Phase onset achieved a significantly lower classification accuracy than the windows extracted from 3-6,6-9, and 12-15 seconds and a highly significantly lower accuracy than the windows from 9-12 seconds (see Figure \ref{fig:part}).

If we restrict the trial-sensitive splitting of the training data further by using only the last 30\% of trials for testing in the BCI-approach, the average classification accuracy drops to 66.38\% $\pm$ 14.5\%. This overall average classification accuracy is significantly better than random and individually, 11 of the 20 participants had a classification accuracy better than random. For the results per person, see the second column (BCI-approach) of Table \ref{tab:overview}.

\subsection{Eye Tracking Classification}
\begin{figure}[h]
    \centering
    \includegraphics[width = 0.7\textwidth]{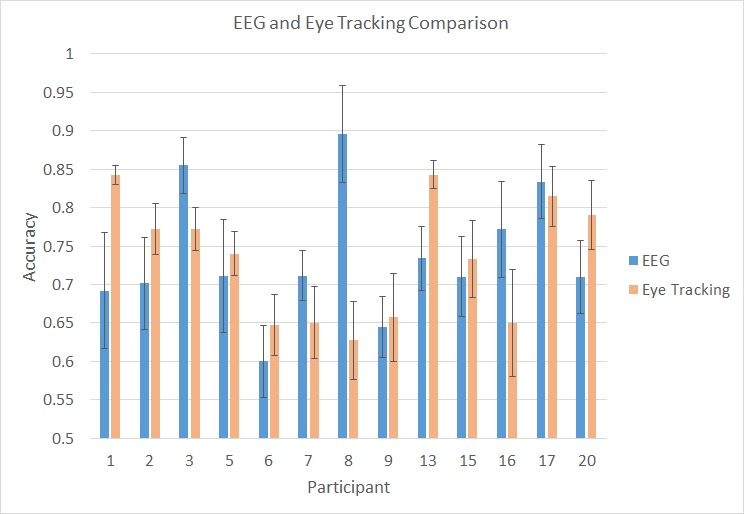}
    \caption{Direct comparison of EEG and eye tracking classification results individually. The barplots show the mean and standard deviation over the 10 trial-sensitive randomized runs per participant. }
    \label{fig:eye}
\end{figure}

Technical problems arose during the recordings of the eye tracking data. For 7 participants, either a successful calibration was not possible because the pupil detection was not stable, or the confidence of the eye tracker decreased dramatically because the lightning conditions changed and again, pupil detection was not possible. This shows that EEG may often be more reliable than ET.\\
Figure \ref{fig:eye} shows a comparison of the 13 participants with full EEG and eye tracking datasets of the study. We compared the mean classification accuracy over 10 classification runs for both datasets for all subjects. The mean accuracy for the eye tracking data was 73.39\% $\pm$ 7.6\% (EEG: 73.63\% $\pm$ 7.99\%). The accuracy difference between the two modalities was not significant (t(12)= 0.0784, p=.9388). For 5 of the 13 participants, the classifier performance was better for the EEG data, and for 8 participants, it was better for the eye tracking data. There was no correlation between the results for eye tracking and EEG (Pearson's r = 0.07).

\begin{figure}
    \centering
    \includegraphics[width = \textwidth]{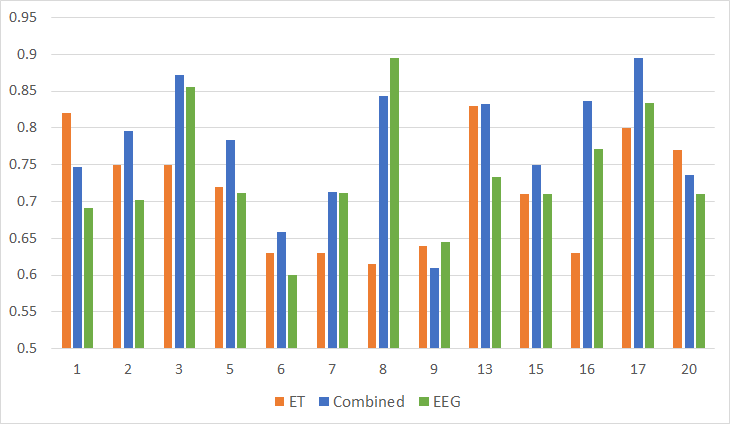}
    \caption{Classification accuracies of the single modalities compared to the combined modalities with a late fusion approach per participant}
    \label{fig:latefusion}
\end{figure}

For the late fusion approach that combined both modalities, the average classification accuracy over all 13 subjects increased to 77.47\% $\pm$ 8\%. This improvement was significant compared to the EEG only result \textit{(t(12)=3.0114, p= 0.0108)} but not compared to the eye tracking result \textit{(t(12)=1.6343, p = 0.1281)}. On average, 71.2\% $\pm$ 10.66\% of the predictions were based on the EEG model. Overall, the results improved for 11 of the 13 participants (See Figure \ref{fig:latefusion})

\subsection{Person-Independent Classification}
For the person-independent EEG classification, the training was performed on the data of 19 subjects and tested on the remaining subject. Following the significance analysis method of \cite{mullerputz}, with $n=200$ and $p=0.5$, the classification accuracy has to be above 56.8\% to be considered better than random. The individual person-independent classification results were better than random for 6 of the 20 participants (see column "Person-Independent" of Table \ref{tab:overview}). The mean accuracy over all participants reached 54\% $\pm$ 6\% which is not significantly better than random.\\

\subsection{Feature Analysis}
\begin{figure}[h]
    \centering
    \includegraphics[width=\textwidth]{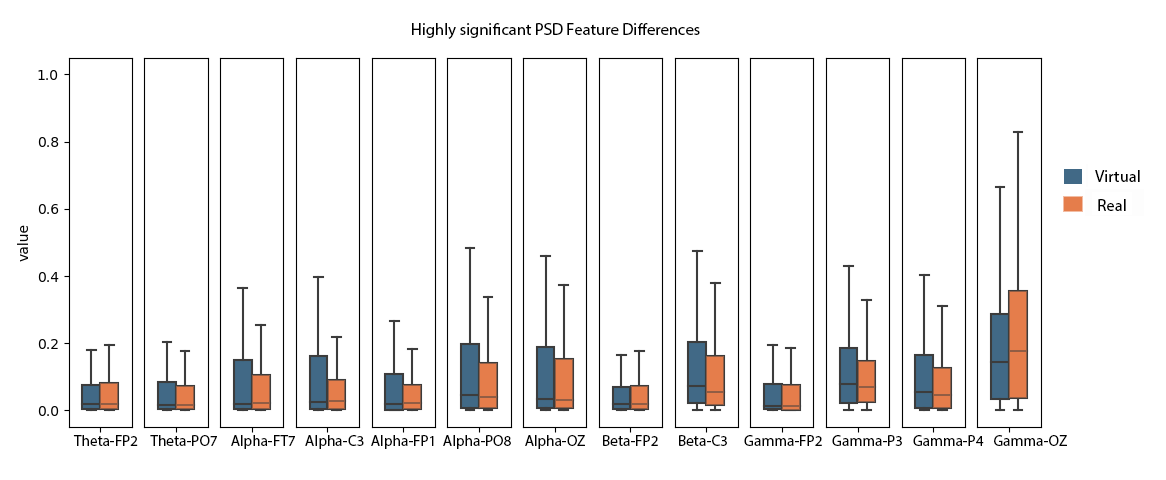}
    \caption{Boxplot visualization of the features that show highly significant ($\alpha<0.001$) differences between the two conditions when pooled over all participants. The data war scaled between 0 and 1 for each participant. The bar shows the mean, the box shows the quartiles of the dataset, and the whiskers extend to show the rest of the distribution, outliers are not included.}
    \label{fig:features}
\end{figure}

To answer the question what the FBCSP-based neural net learned, different approaches and visualizations were tested. Firstly, we decided to compare frequency band features of the two task conditions independent of the models. We tested for significant differences if the features are pooled over all subjects. With a significance level of $\alpha<0.001$ for a highly significant difference, 13 features were selected. A visualization for the 13 features is shown in Figure \ref{fig:features}. For the theta-band, the differences between conditions were significant for FP2 \textit{(t(3939)=-4.356)} and for PO7 \textit{(t(3939)=-3.749)}. Most significant differences over all subjects were found for the alpha-band, namely for FT7 \textit{(t(3939)=5.0524)}, C3 \textit{(t(3939)=8.741)}, FP1  \textit{(t(3939)=6.694)}, PO8 \textit{(t(3939)=5.365)}, and Oz  \textit{(t(3939)=4.881)}. Highly significant differences in the beta-band were present in the locations of FP2 \textit{(t(3939)=-3.606)} and C3  \textit{(t(3939)=4.679)}. For the gamma-band, the selected electrode positions were FP2\textit{ (t(3939)=-3.329)}, P3 \textit{(t(3939)=3.419)}, P4 \textit{(t(3939)=4.490)}, and Oz \textit{(t(3939)=-6.711)}. Overall, the selected electrodes were mainly located in parietal and occipetal regions of the brain, which is in accordance with the results from \cite{Liang} on visual attention. \\
We also visualized and analyzed what the neural net had learned. However, no common pattern was shared across participants or across person-independent models and the results were not visible in this approach. 

\section{Discussion}
In this work, we used machine learning to classify attentional focus on real and virtual objects in Augmented Reality. To the best of our knowledge, this is the first work on the subject. We implemented an adjusted pairs game with real and virtual cards and classified 3-second windows of EEG and eye tracking data.\\

Our first hypothesis was that we would be able to predict the real and virtual targets of attention with an accuracy better than random for person-dependent classifiers. Even after excluding the time effect in the data, the average accuracy over all participants was still significantly higher than chance and individually, 17 of 20 subjects reached a performance significantly better than chance. This reliable prediction proves our hypothesis.\\
One drawback of BCI systems is that they are not very robust and rarely reach 100\% accuracy. Usually, their results worsen with time as movements and environmental factors influence the accuracy \citep{lotte2015}. Additionally, it has been claimed that 20\% of the participants in a BCI experiment are unable to achieve reliable results because of "BCI-Illiteracy" \citep{allison}. The reasons for this effect have not yet been determined but could be due to the high interpersonal differences between brain activation patterns \citep{edlinger}.\\
Considering the short time windows of 3 seconds and the sparse positioning of the electrodes, the results are satisfying. We assume that the classification accuracy can be improved by increasing the density of the EEG or optimizing the current placement of the electrodes. Moreover, longer data windows could improve the performance of the classifier but with the goal of real-time classification in mind, this analysis was not performed. Further tuning of the setup and preprocessing, as well as a more specialized classification process, could significantly improve the classification accuracies for each individual participant. \\

Since the distinction is possible based on the EEG data, the question remains what exactly leads to the differences in neural activity patterns for the two conditions. The individual and person-independent feature analyses showed many variances between participants. Possible explanations for distinguishable cognitive user states in this task include aspects of workload, memory, or visual input properties. Depending on the participant, the main perceived differences between the two task conditions might vary. The frequency band analysis sheds some light on the possible distinguishing features: Most features with highly significant differences over all participants were located in parietal and occipital regions of the brain. The occipital cortex contains the visual cortex, which is the primary region for the processing of visual information, while the parietal cortex is activated through sensations and perception and plays an important role for sensory integration, especially for visual input \citep{ungerleider, buracas, herrmann}.\\
The two frequency bands that showed the most significant differences were the gamma and the alpha band. In \cite{neuner}, it was shown that gamma-band activity, especially in occipital regions, was synchronous to attention. \cite{jensen} state that the gamma band activity is present when information is presented in a more complicated way. According to \cite{min}, alpha-band activity in the parietal and occipital regions is reduced for retained attention to a bright stimulus.\\
These initial clues about the differences should be used in further studies on the topic for both setup and additional examination.\\

Compared to the EEG recordings, the eye tracking recordings had a worse quality and more technical difficulties. These problems could be solved in the future with a built-in eye tracker or a different setup. Our second hypothesis was that the eye tracking classification would not significantly outperform the EEG data classification. For our features and classification approaches, this proved to be true. It was shown that the classification accuracy of EEG and eye tracking data did not correlate and that in many cases either the classification of the EEG or of the eye tracking data did lead to very high classification accuracies. This suggests that a combination of the two modalities (either in an early or a late fusion approach) would be beneficial for the performance of a classifier. More work would have to be put into the extraction of sophisticated features for this task and a suitable decision function or feature combination could combine the advantages of both approaches. However, finding an optimized classifier was not in the scope of this work.\\

Our goal was to perform a first study that tests whether attention on real and virtual objects in AR is represented differently in the brain to an extent that makes machine learning-based classification possible. Even an initial attempt at a person-independent EEG-based classification showed promising results. Thus, we conclude that further research on this topic will attain interesting and useful results for the improvement of Augmented Reality devices and applications.

\subsection{Future Work}
Following the positive results from this study, the next steps will include the implementation of other scenarios to test whether the results were task-dependent. The scenarios will be less static and controlled while improving the setup based on the newly gained knowledge from this study. We know now that the classification is possible and we found plausible features that differ between the conditions for all participants. One question that remained unanswered is whether the classification accuracy decreases for highly experienced users. The participants in this study had some experience but not on a level where we would expect strong differences in the perception. One idea is to compare a group of very experienced AR users with inexperienced users.\\
In general, the perceived differences between users in AR are an interesting topic and the mentioned aspects of workload, memory or consciousness of the perception of virtual information are worth further studying.\\
The improved setup for the next studies will also include adjusted classification processes. The combination of EEG and eye tracking into a multimodal classifier seems promising and even gaze point information could be included as clues for specific applications. Additionally, Person-independent classification can be improved and eye tracking data will be included for this. A completely different approach to the classification of attention in this context would be the analysis of the SSVEP, based on the display frequency of the Augmented Reality device. SSVEP detection with a flickering frequency has been done before \citep{60hz}.\\
The overall goal is an application that profits from the real-time classification of attention on real and virtual objects in AR.

\section*{Conflict of Interest Statement}

The authors declare that the research was conducted in the absence of any commercial or financial relationships that could be construed as a potential conflict of interest.

\section*{Author Contributions}

The study was planned by LV, LS, and FP. The experiment implementation and data collection was completed by LS. LV performed the data analysis and wrote the paper. FP supervised the process. 

\section*{Funding}
This research was funded by the "Zentrale Forschungsförderung" of the University of Bremen in the context of the project "Attention-driven Interaction Systems in Augmented Reality". Open access was supported by the Open Access Initiative of the University of Bremen and the DFG.

\section*{Data Availability Statement}
The datasets generated and analyzed for this study, as well as the task,  can be found online. The link will be announced.
\bibliography{main}

\begin{thebibliography}{37}
\providecommand{\natexlab}[1]{#1}
\expandafter\ifx\csname urlstyle\endcsname\relax
  \providecommand{\doi}[1]{doi:\discretionary{}{}{}#1}\else
  \providecommand{\doi}{doi:\discretionary{}{}{}\begingroup
  \urlstyle{rm}\Url}\fi
\providecommand{\selectlanguage}[1]{\relax}
\providecommand{\bibAnnoteFile}[1]{%
  \IfFileExists{#1}{\begin{quotation}\noindent\textsc{Key:} #1\\
  \textsc{Annotation:}\ \input{#1}\end{quotation}}{}}
\providecommand{\bibAnnote}[2]{%
  \begin{quotation}\noindent\textsc{Key:} #1\\
  \textsc{Annotation:}\ #2\end{quotation}}

\bibitem[{Allison and Neuper(2010)}]{allison}
Allison, B. and Neuper, C. (2010).
\newblock \emph{Could anyone use a BCI?}
\newblock 35--54.
\newblock \doi{10.1007/978-1-84996-272-8_3}
\bibAnnoteFile{allison}

\bibitem[{Azuma(1997)}]{Azuma}
Azuma, R.~T. (1997).
\newblock A survey of augmented reality.
\newblock \emph{Presence: Teleoper. Virtual Environ.} 6, 355–385.
\newblock \doi{10.1162/pres.1997.6.4.355}
\bibAnnoteFile{Azuma}

\bibitem[{{Barresi} et~al.(2015){Barresi}, {Olivieri}, {Caldwell}, and
  {Mattos}}]{barresi2015}
{Barresi}, G., {Olivieri}, E., {Caldwell}, D.~G., and {Mattos}, L.~S. (2015).
\newblock Brain-controlled ar feedback design for user's training in surgical
  hri.
\newblock In \emph{2015 IEEE International Conference on Systems, Man, and
  Cybernetics}. 1116--1121.
\newblock \doi{10.1109/SMC.2015.200}
\bibAnnoteFile{barresi2015}

\bibitem[{Buracas and Boynton(2007)}]{buracas}
Buracas, G. and Boynton, G. (2007).
\newblock The effect of spatial attention on contrast response functions in
  human visual cortex.
\newblock \emph{The Journal of neuroscience : the official journal of the
  Society for Neuroscience} 27, 93--7.
\newblock \doi{10.1523/JNEUROSCI.3162-06.2007}
\bibAnnoteFile{buracas}

\bibitem[{Busch and VanRullen(2010)}]{busch}
Busch, N. and VanRullen, R. (2010).
\newblock Spontaneous eeg oscillations reveal periodic sampling of visual
  attention.
\newblock \emph{Proceedings of the National Academy of Sciences of the United
  States of America} 107, 16048--53.
\newblock \doi{10.1073/pnas.1004801107}
\bibAnnoteFile{busch}

\bibitem[{Day(1964)}]{day1964}
Day, M.~E. (1964).
\newblock An eye movement phenomenon relating to attention, thought and
  anxiety.
\newblock \emph{Perceptual and Motor Skills} 19, 443--446.
\newblock \doi{10.2466/pms.1964.19.2.443}.
\newblock PMID: 14214714
\bibAnnoteFile{day1964}

\bibitem[{Edlinger et~al.(2015)Edlinger, Allison, and Guger}]{edlinger}
Edlinger, G., Allison, B.~Z., and Guger, C. (2015).
\newblock \emph{How Many People Can Use a BCI System?} (Tokyo: Springer Japan).
\newblock 33--66.
\newblock \doi{10.1007/978-4-431-55037-2_3}
\bibAnnoteFile{edlinger}

\bibitem[{Florea et~al.(2013)Florea, Florea, Vranceanu, and
  Vertan}]{florea2013}
Florea, L., Florea, C., Vranceanu, R., and Vertan, C. (2013).
\newblock Can your eyes tell me how you think? a gaze directed estimation of
  the mental activity.
\newblock 60.1--60.11.
\newblock \doi{10.5244/C.27.60}
\bibAnnoteFile{florea2013}

\bibitem[{Guger et~al.(2012)Guger, Allison, Grosswindhager, Prückl,
  Hintermüller, Kapeller et~al.}]{SSVEP}
Guger, C., Allison, B., Grosswindhager, B., Prückl, R., Hintermüller, C.,
  Kapeller, C., et~al. (2012).
\newblock How many people could use an ssvep bci?
\newblock \emph{Frontiers in Neuroscience} 6, 169.
\newblock \doi{10.3389/fnins.2012.00169}
\bibAnnoteFile{SSVEP}

\bibitem[{Herrmann(2001)}]{herrmann}
Herrmann, C. (2001).
\newblock Human eeg responses to 1–100 hz flicker: Resonance phenomena in
  visual cortex and their potential correlation to cognitive phenomena.
\newblock \emph{Experimental brain research. Experimentelle Hirnforschung.
  Expérimentation cérébrale} 137, 346--53.
\newblock \doi{10.1007/s0022101}
\bibAnnoteFile{herrmann}

\bibitem[{Huang et~al.(2019)Huang, Li, Ngai, Leong, and Bulling}]{Huang}
Huang, M.~X., Li, J., Ngai, G., Leong, H.~V., and Bulling, A. (2019).
\newblock Moment-to-moment detection of internal thought during video viewing
  from eye vergence behavior.
\newblock In \emph{Proceedings of the 27th ACM International Conference on
  Multimedia} (New York, NY, USA: Association for Computing Machinery), MM
  ’19, 2254–2262.
\newblock \doi{10.1145/3343031.3350573}
\bibAnnoteFile{Huang}

\bibitem[{Hutton and Ettinger(2006)}]{hutton2006}
Hutton, S. and Ettinger, U. (2006).
\newblock Hutton sb, ettinger u. the antisaccade task as a research tool in
  psychopathology: a critical review. psychophysiology 43: 302-313.
\newblock \emph{Psychophysiology} 43, 302--13.
\newblock \doi{10.1111/j.1469-8986.2006.00403.x}
\bibAnnoteFile{hutton2006}

\bibitem[{Jensen et~al.(2007)Jensen, Kaiser, and Lachaux}]{jensen}
Jensen, O., Kaiser, J., and Lachaux, J.-P. (2007).
\newblock Human gamma-frequency oscillations associated with attention and
  memory.
\newblock \emph{Trends in neurosciences} 30, 317--24.
\newblock \doi{10.1016/j.tins.2007.05.001}
\bibAnnoteFile{jensen}

\bibitem[{Jiang et~al.(2019)Jiang, Wang, Pei, and Chen}]{60hz}
Jiang, L., Wang, Y., Pei, W., and Chen, H. (2019).
\newblock A four-class phase-coded ssvep bci at 60hz using refresh rate.
\newblock vol. 2019, 6331--6334.
\newblock \doi{10.1109/EMBC.2019.8857326}
\bibAnnoteFile{60hz}

\bibitem[{Kinsbourne(2013)}]{Kinsbourne}
Kinsbourne, M. (2013).
\newblock Neuropsychology of attention \doi{10.1016/B978-0-08-092668-1.50011-9}
\bibAnnoteFile{Kinsbourne}

\bibitem[{Krigolson et~al.(2017)Krigolson, Williams, and Colino}]{krigolson}
Krigolson, O., Williams, C., and Colino, F. (2017).
\newblock Using portable eeg to assess human visual attention.
\newblock 56--65.
\newblock \doi{10.1007/978-3-319-58628-1_5}
\bibAnnoteFile{krigolson}

\bibitem[{Liang et~al.(2017)Liang, Lin, Yao, Chang, Liu, and Chen}]{Liang}
Liang, C., Lin, C.-T., Yao, S.-N., Chang, W.-S., Liu, Y.-C., and Chen, S.-A.
  (2017).
\newblock Visual attention and association: An electroencephalography study in
  expert designers.
\newblock \emph{Design Studies} 48, 76 -- 95.
\newblock \doi{https://doi.org/10.1016/j.destud.2016.11.002}
\bibAnnoteFile{Liang}

\bibitem[{Liu et~al.(2013)Liu, Chiang, and Chu}]{liu2013}
Liu, N.-H., Chiang, C.-Y., and Chu, H.-C. (2013).
\newblock Recognizing the degree of human attention using eeg signals from
  mobile sensors.
\newblock \emph{Sensors (Basel, Switzerland)} 13, 10273--86.
\newblock \doi{10.3390/s130810273}
\bibAnnoteFile{liu2013}

\bibitem[{Lotte et~al.(2015)Lotte, Bougrain, and Clerc}]{lotte2015}
Lotte, F., Bougrain, L., and Clerc, M. (2015).
\newblock \emph{Electroencephalography (EEG)-Based Brain–Computer Interfaces}
  (American Cancer Society).
\newblock 1--20.
\newblock \doi{10.1002/047134608X.W8278}
\bibAnnoteFile{lotte2015}

\bibitem[{Min et~al.(2013)Min, Jung, Kim, and Park}]{min}
Min, B.-K., Jung, Y.-C., Kim, E., and Park, J.~Y. (2013).
\newblock Bright illumination reduces parietal eeg alpha activity during a
  sustained attention task.
\newblock \emph{Brain Research} 1538, 83 -- 92.
\newblock \doi{https://doi.org/10.1016/j.brainres.2013.09.031}
\bibAnnoteFile{min}

\bibitem[{Myrden and Chau(2015)}]{myrden}
Myrden, A. and Chau, T. (2015).
\newblock Effects of user mental state on eeg-bci performance.
\newblock \emph{Frontiers in Human Neuroscience} 9, 308.
\newblock \doi{10.3389/fnhum.2015.00308}
\bibAnnoteFile{myrden}

\bibitem[{Müller-Putz et~al.(2008)Müller-Putz, Scherer, Brunner, Leeb, and
  Pfurtscheller}]{mullerputz}
Müller-Putz, G., Scherer, R., Brunner, C., Leeb, R., and Pfurtscheller, G.
  (2008).
\newblock Better than random? a closer look on bci results.
\newblock \emph{International Journal of Bioelektromagnetism} 10, 52--55
\bibAnnoteFile{mullerputz}

\bibitem[{Neuner et~al.(2014)Neuner, Arrubla, Werner, Hitz, Boers, Kawohl
  et~al.}]{neuner}
Neuner, I., Arrubla, J., Werner, C., Hitz, K., Boers, F., Kawohl, W., et~al.
  (2014).
\newblock The default mode network and eeg regional spectral power: A
  simultaneous fmri-eeg study.
\newblock \emph{PloS one} 9, e88214.
\newblock \doi{10.1371/journal.pone.0088214}
\bibAnnoteFile{neuner}

\bibitem[{Pedregosa et~al.(2012)Pedregosa, Varoquaux, Gramfort, Michel,
  Thirion, Grisel et~al.}]{scikit}
[Dataset] Pedregosa, F., Varoquaux, G., Gramfort, A., Michel, V., Thirion, B.,
  Grisel, O., et~al. (2012).
\newblock Scikit-learn: Machine learning in python
\bibAnnoteFile{scikit}

\bibitem[{Putze et~al.(2013)Putze, Hild, Kärgel, Herff, Redmann, Beyerer
  et~al.}]{putze2013}
Putze, F., Hild, J., Kärgel, R., Herff, C., Redmann, A., Beyerer, J., et~al.
  (2013).
\newblock Locating user attention using eye tracking and eeg for
  spatio-temporal event selection.
\newblock 129--136.
\newblock \doi{10.1145/2449396.2449415}
\bibAnnoteFile{putze2013}

\bibitem[{Putze et~al.(2019)Putze, Weiß, Vortmann, and Schultz}]{putze2019}
Putze, F., Weiß, D., Vortmann, L.-M., and Schultz, T. (2019).
\newblock Augmented reality interface for smart home control using ssvep-bci
  and eye gaze.
\newblock 2812--2817.
\newblock \doi{10.1109/SMC.2019.8914390}
\bibAnnoteFile{putze2019}

\bibitem[{Rensink(2015)}]{Rensink}
Rensink, R. (2015).
\newblock \emph{A Function-Centered Taxonomy of Visual Attention}.
\newblock 347--375.
\newblock \doi{10.1093/acprof:oso/9780198712718.003.0013}
\bibAnnoteFile{Rensink}

\bibitem[{Sauseng et~al.(2006)Sauseng, Klimesch, Stadler, Schabus, Doppelmayr,
  Hanslmayr et~al.}]{Sauseng}
Sauseng, P., Klimesch, W., Stadler, W., Schabus, M., Doppelmayr, M., Hanslmayr,
  S., et~al. (2006).
\newblock A shift of visual spatial attention is selectively associated with
  human eeg alpha activity.
\newblock \emph{The European journal of neuroscience} 22, 2917--26.
\newblock \doi{10.1111/j.1460-9568.2005.04482.x}
\bibAnnoteFile{Sauseng}

\bibitem[{Schirrmeister et~al.(2017)Schirrmeister, Springenberg, Fiederer,
  Glasstetter, Eggensperger, Tangermann et~al.}]{Schirrmeister}
Schirrmeister, R.~T., Springenberg, J.~T., Fiederer, L. D.~J., Glasstetter, M.,
  Eggensperger, K., Tangermann, M., et~al. (2017).
\newblock Deep learning with convolutional neural networks for eeg decoding and
  visualization.
\newblock \emph{Human Brain Mapping} 38, 5391--5420.
\newblock \doi{10.1002/hbm.23730}
\bibAnnoteFile{Schirrmeister}

\bibitem[{Si-Mohammed et~al.(2017)Si-Mohammed, Argelaguet, Casiez, Roussel, and
  Lécuyer}]{simohammed2017}
Si-Mohammed, H., Argelaguet, F., Casiez, G., Roussel, N., and Lécuyer, A.
  (2017).
\newblock Brain-computer interfaces and augmented reality: A state of the art.
\newblock \doi{10.3217/978-3-85125-533-1-82}
\bibAnnoteFile{simohammed2017}

\bibitem[{Sibert et~al.(2000)Sibert, Jacob, and Templeman}]{silbert2000}
Sibert, L., Jacob, R., and Templeman, J. (2000).
\newblock Evaluation and analysis of eye gaze interaction
\bibAnnoteFile{silbert2000}

\bibitem[{Ungerleider and G.(2000)}]{ungerleider}
Ungerleider, S.~K. and G., L. (2000).
\newblock Mechanisms of visual attention in the human cortex.
\newblock \emph{Annual Review of Neuroscience} 23, 315--341.
\newblock \doi{10.1146/annurev.neuro.23.1.315}.
\newblock PMID: 10845067
\bibAnnoteFile{ungerleider}

\bibitem[{Vortmann et~al.(2019{\natexlab{a}})Vortmann, Kroll, and
  Putze}]{vortmann2019}
Vortmann, L.-M., Kroll, F., and Putze, F. (2019{\natexlab{a}}).
\newblock Eeg-based classification of internally- and externally-directed
  attention in an augmented reality paradigm.
\newblock \emph{Frontiers in Human Neuroscience} 13, 348.
\newblock \doi{10.3389/fnhum.2019.00348}
\bibAnnoteFile{vortmann2019}

\bibitem[{Vortmann and Putze(2020)}]{vortmannchi}
Vortmann, L.-M. and Putze, F. (2020).
\newblock Attention-aware brain computer interface to avoid distractions in
  augmented reality.
\newblock In \emph{Extended Abstracts of the 2020 CHI Conference on Human
  Factors in Computing Systems} (New York, NY, USA: Association for Computing
  Machinery), CHI EA ’20, 1–8.
\newblock \doi{10.1145/3334480.3382889}
\bibAnnoteFile{vortmannchi}

\bibitem[{Vortmann et~al.(2019{\natexlab{b}})Vortmann, Schult, Benedek,
  Annerer-Walcher, and Putze}]{vortmannicmi}
Vortmann, L.-M., Schult, M., Benedek, M., Annerer-Walcher, S., and Putze, F.
  (2019{\natexlab{b}}).
\newblock Real-time multimodal classification of internal and external
  attention.
\newblock 1--7.
\newblock \doi{10.1145/3351529.3360658}
\bibAnnoteFile{vortmannicmi}

\bibitem[{Zander and Kothe(2011)}]{zander}
Zander, T. and Kothe, C. (2011).
\newblock Towards passive brain–computer interfaces: applying
  brain–computer interface technology to human-machine systems in general.
\newblock \emph{Journal of neural engineering} 8, 025005.
\newblock \doi{10.1088/1741-2560/8/2/025005}
\bibAnnoteFile{zander}

\bibitem[{Zhang and Yu(2010)}]{zhang}
Zhang, C. and Yu, X. (2010).
\newblock Estimating mental fatigue based on electroencephalogram and heart
  rate variability.
\newblock \emph{Pol J Med Phys Eng PL ISSN} 16, 67--84.
\newblock \doi{10.2478/v10013-010-0007-7}
\bibAnnoteFile{zhang}

\end{thebibliography}

\end{document}